**Review: Graphene-based biosensor for Viral Detection**

*Amine El Moutaouakil*, Suhada Poovathy, Mohamed Belmoubarik and Weng Kung Peng*

Prof. A. El Moutaouakil*, S. Poovathy

Department of Electrical Engineering, College of Engineering, UAE University, Al Ain, UAE
E-mail: a.elmoutaouakil@uaeu.ac.ae

Dr. M. Belmoubarik, Dr. W. K. Peng
International Iberian Nanotechnology Laboratory, Portugal

Viral infections are among the main reasons for serious pandemics and contagious infections; hence, they cause thousands of fatalities and economic losses annually. In the case of COVID-19, world economies have shut down for months, and physical distancing along with drastic changes in the social behavior of many humans has generated many issues for all countries. Thus, a rapid, low-cost, and sensitive viral detection method is critical to upgrade the living standards of humans while exploiting biomedicine, environmental science, bioresearch, and biosecurity. The emergence of various carbon-based nanomaterials such as carbon nanotubes, graphene, and carbon nanoparticles provided a great possibility for researchers to develop a new and wide variety of biosensors. In particular, graphene has become a promising tool for biosensor fabrication owing to its many interesting properties, such as its exceptional conductivity, ultrahigh electron mobility, and excellent thermal conductivity. This paper provides an overall perspective of graphene-based biosensors and a critical review of the recent advances in graphene-based biosensors that are used to detect different types of viruses, such as Ebola, Zika, and influenza.

**1. Introduction**

There is a remarkable rise in the number of infectious diseases, which increases the urge for a simple and quick mechanism for the early recognition of pathogenic viruses. Numerous diagnostic and detection tests are currently available, but infectious diseases still pose an omnipresent threat to public health globally, especially in many developing countries. Many cases of infections went undiagnosed or diagnosed at later stages due to the limitation of accessing to the right (sensitive) diagnostic tools. The World Health Organization formulated a step-wise guide, "Affordable, Sensitive, Specific, User-friendly, Rapid and robust, Equipment-free and Deliverable to end-users" or ASSURED in 2017,[1] to help healthcare professionals in selecting, implementing, and monitoring appropriate diagnostic tests in resource-constrained environments.[2,3]

The early and rapid determination of pathogenic and infectious agents, such as bacteria, viruses, and parasites, is crucial to select the correct control measures regarding the containment and confinement of diseases and the use of antimicrobials and vaccines, especially when combating emerging infectious diseases and biothreat agents. Generally, viral cultivation is the benchmark for virus detection, as it is a laboratory procedure where samples are kept with certain cell types in which viruses are tested for their ability to infect other cells. This method is accurate and sensitive, but it usually takes up to several days to get the results.[2] The currently available clinical laboratory methods such as immunology-based methods, polymerase chain reaction (PCR), electrochemical methods, and enzyme-linked immunosorbent assay (ELISA) need high-cost reagents, specific laboratory conditions, well-trained staff, highly accurate instruments, and calibration methods so as to provide reliable and reproducible sensitivity.[4] Table 1 summarizes the main differences between these methods.

Biosensor technology offers low-cost and miniature detection platform apart from providing reliable results in much shorter times in comparison with conventional detection methods, while the improvements of selectivity, performance, and stability is an on-going subject of research.[5,6] Interestingly, integrating the advances in nanomaterials with biosensor technology resulted in nano-biosensors, which are emerging tools that can carry out bacterial or viral detection and identification.[7] Nano-biosensors offer many key advantages, including improvements in sensitivity and specificity, nano-scale integration, and real-time detection, and they also enable the use of novel, diverse, and label-free sensing mechanisms.[6,8,9] Among the different available nanomaterials, graphene is a promising candidate for fabricating state-of-the-art nano-scale sensors and biosensors due to its unique mechanical, physical, electrical, and optical properties.[10–14] Extensive researchers included graphene nanomaterials to develop novel label-free, real-time, and ultrasensitive sensors to monitor various diseases (Figure 1).[5,7,14–17]

This paper briefly describes the architecture of graphene-based biosensors, and it reviews the recent advances in the field. Section 1 gives a brief outline of biosensors and the incorporation of graphene into a standard biosensor.[11,12] The following sections discuss various virus sensors that are based on simple graphene[7,18–22] and hybrid/advanced graphene structures.[23–29] Subsequently, we discuss the expected future advancements in the use of graphene-based biosensors and the current challenges.

## 2. Theoretical Model

A biosensor is defined as an analytical device comprising a biological element in close contact with a physiochemical transducer that generates an electronic signal proportional to the analyte or concentration of a target substance in a sample. Usually, it consists of a biological recognition component, a sensor element for signal acquisition, and a feeding signal processing and display unit (Figure 2).[7,8,11] Biosensors have a broad variety of uses in the biomedical and healthcare fields, including the detection of pathogenic microorganisms,[7] medical implants, the assistance in drug delivery,[5] and the food and agricultural industries.[16] Also, label-free biosensors with improved sensitivity are essential for consumer genetics,[15] the diagnosis and treatment of cancer,[30–32] the control of serious pandemics, and the drug development field.[11,33] Different types of biosensors are available depending on the types of recognition elements, immobilization techniques, and transduction principles.

## 3. Integration of Graphene within the Biosensor

Graphene is a crystalline allotrope of carbon with a single layer of sp2-hybridized carbon atoms with a carbon–carbon bond length of 0.142 nm in a hyper-dimensional (1D, 2D, or 3D) structur.[11] It has many unique properties, including a large surface area of 2630 $m^2.g^{-1}$. The electrical conductivity of synthetic and natural graphene exceeds 1000 siemens per meter, and its measured thermal conductivity is in the range of 3000–5000 $W.m^{-1}.K^{-1}$.[6] It also has high mechanical strength, a Young's modulus of order ~1.0 TPa, and a tunable bandgap and tensile strength, making it a fascinating element for next-generation biosensors.[14]

Several techniques are used for the preparation and synthesis of graphene on different substrates, including exfoliation, the epitaxial growth on silicon carbide, the Hummers method, chemical vapor deposition (CVD), the reduction of graphite oxide, and the unzipping of multiwalled carbon nanotubes. Also, different carbon-based materials such as pristine graphene, graphene oxide (GO), reduced GO (rGO), and graphene quantum dot (Figure 3) can be synthesized using these methods. Among these, the most efficient, inexpensive, and common technique for bulk scale preparation is the CVD technique, which entails the depositing of gaseous reactants onto a substrate in a gas chamber, where the quality of the produced graphene can be controlled using the varying pressure and temperature feature of the technique.[34,35]

The biomolecules, such as antibodies, enzymes, and DNA, can be easily incorporated into the graphene's detection scheme (Figure 4), which is why extensive research on the development of graphene biosensors for various applications has been conducted. The immobilization of specific analyte by adsorption, covalent binding, entrapment, or membrane confinement onto the bioreceptor results in a change in the electrical properties, which in turn results in a measurable electrical response. Hence, the presence or concentration of the analyte of interest can be identified.[12,13]

Furthermore, the large transducing area of graphene provides a large catalytic action. This atom-thin layer yet strong and biocompatible material shows a high electron transfer rate; hence, it can be used for highly efficient biosensor applications.[10] Graphene can also be combined with other metal compounds such as gold or other nanoparticles to obtain more specific and sensitive sensors with lower detection limits, which can open the way to better sensing capabilities, and wider applications. Moreover, it exhibits high optical transparency to visible-light wavelengths, so it can be used in optical-detection biosensors and in bio-tissue observation.[36] Recently, graphene was used in the fabrication of wearable sensors and medical implants because of its ideal bio-integrating properties with human skin.[17,37–40]

**4. Architecture of Graphene-based Virus Biosensors**

There are mainly four approaches in graphene biosensors:

**4-1.    Electron Transfer Materials**

The graphene's large surface area exposes all the carbon atoms to the analyte; hence, it can be used as an electron transfer material for electrochemical biosensors. Direct electrical communication is facilitated by the physical binding to the analyte, and the electrons in graphene have very high mobility, thus ensuring the movement of electrons in the sensors.[41]

**4-2.    Impedance Materials**

The basic graphene atom forms three σ-bonds with each of its three neighbors and one π-bond that forms π–π stacking interactions. Hence, other π-conjugated molecules can easily be attached to the electrode. So the graphene's surface can be easily modified without upsetting its electrical properties.[42]

**4-3.    Photon and Phonon Transfer**

The unique nature of the phonon transport properties (acoustic phonons and atomic phonons) in graphene made its application in the plasmonic fields and the coupling electromagnetic radiation in the terahertz frequency possible. Electrochemiluminescence and fluorescence are the main means for photon and phonon transfer, so they are used with surface plasmon polaritons on metal films to improve the performance of biosensors. Surface plasmons are coherent electrons that propagate along the interface between a metal and a dielectric, and these waves quantize the longitudinal plasma oscillations of free electrons near the metallic surface. Graphene-based surface plasmon resonance (SPR) sensors are developed for direct and indirect pathogen detection with improved sensitivity in comparison with the conventional methods.[36,43] The sensitivity and strength of optical detection are enhanced by layering graphene on the top of standard metals such as gold and silver. Thus, the integration of graphene into optical sensors and other types of sensors results in the development of ultrasensitive biosensors for the detection of protein molecules at the molecular level. Hence, it makes the diagnosis and treatment of a variety of dangerous diseases, such as cancer and Alzheimer's disease, more efficient.[44]

**4-4. Graphene Field-Effect Transistor (GFET)**

The thin silicon channel in field-effect transistors (FETs) is replaced with a thin graphene film to make GFETs so as to enable the easy functionalization and binding of receptor molecules to biosensors and provide a thinner and more sensitive channel region (Figure 5).[45] The surface of the GFET channel can be functionalized by binding receptors such as proteins, bio-compounds, and DNA molecules to make sensors for various applications.[46–49] Graphene shows a high surface-to-volume ratio, which enables the identification of the smallest concentrations of analytes, where the binding of even a single analyte creates a change in the electric field and properties of the system. GFET biosensors are thus useful for the ultrasensitive, rapid, label-free detection of pathogenic molecules, such as viruses and bacteria.[37] They are easily used, fast, and ultrasensitive, and they can be selected for various analytes by attaching specific probes on the graphene channel.[47] Bio-FET can be classified on the basis of the bio-recognition element that is used for the sensing purposes of analytes, such as enzyme FET, immunologically modified FET, DNA-modified FET, gene-modified FET, and "beetle/chip" FET.[50]

For example, Saeed S. Hashwan et al. showed the electrical performance of GO-based FETs. They explained a step-by-step procedure in the GO preparation and device fabrication. The structural properties were analyzed using photoluminescence, which depends on the π–π interaction in the $sp^2$ domain. The electrical behavior was analyzed by plotting a current–voltage graph for three layers of GO, and the authors concluded it as a promising candidate for the biosensor field.[18]

**5. Graphene-based Virus Biosensor**

Nihar Mohanty et al. utilized chemically modified graphene to develop a single bacterium detection device by immobilizing GO sheets on silica substrates with pre-deposited or post-deposited gold electrode.[48] The developed sensor exhibited highly sensitive detection, where the detection of single cells showed a remarkable increase in the device's conductivity, and the authors recommended these biosensors for developing next-generation devices.

Savannah Afsahi et al. created a cost-effective and ultra-specific graphene biosensor by immobilizing a monoclonal antibody (anti-Zika NS1) to detect the Zika virus (Figure 6).[49] They fabricated the biosensor using commercially available graphene and standard procedures such as photolithography and CVD on a copper surface. Then, it was connected to an electronic reader where all the control parameters and data were performed by a PC. When the target was immobilized onto the channel, a significant change in the channel current and gate capacitance (20% increase) was observed.

Fei Liu et al. developed an electrochemical biosensor using graphene for the detection of rotavirus. A graphene film synthesized from a GO film was utilized in the sensor fabrication.[50] The electrode surface was immobilized with pyrene derivatives and antibodies of the rotavirus. In the proposal, when the target binds to the antibodies, an anodic and cathodic current peak takes place, which can be observed using the cyclic voltammetry process. This method provides a sensitivity of the order of 30.7%, but the prepared GO film exhibited irregular edges and shapes with different sizes, which resulted in low reproducibility.

In 2013, the same research team presented a modified and more sensitive sensing model using rGO FET. This practically overcame the low reproducibility and the other setbacks in their previous work. The antibody was covalently linked to the FET, and the sensitivity was analyzed with different concentrations of the target rotavirus. Each binding and different concentration resulted in a change in the resistance of the measuring device; hence, it was suggested for highly sensitive pathogen detection.[20]

Mudita Pant et al. synthesized another GO biosensor for the sensitive analysis of rotavirus. Here, the crystal surface of the gold microelectrodes was coated with graphene by using the drop-casting method.[47] To illustrate, they fabricated the FET biosensor by simply drop-casting 100 μL of an rGO solution (0.5 mg.mL$^{-1}$) in between the pair of the gold electrodes over a hot plate at a temperature of 70 °C. To enhance the specificity, the attached antibody to the graphene-coated FET crystal was further subjected to bovine serum albumin to block the unused attaching sites so that only the antibody can be attached to the antigen. To examine the sensitivity of the fabricated device, it was exposed to varying concentrations of the target antigen (rotavirus) in the PBS buffer (pH 7.4), and it provided a label-free, rapid, and more specific technique of rotavirus detection.

Yinxi Huang et al. grew graphene using the CVD method, and they characterized it with atomic force microscopy and Raman spectroscopy.[21] The transistor was fabricated by immobilizing anti-*Escherichia coli* antibodies to the graphene film. When the specific bacteria (*E. coli*) adheres to the device, an increase in conductance with respect to the increase in the *E. coli* concentration is observed. They then ensured high sensitivity and specificity by analyzing different bacteria in different pH values and concentrations of analytes and achieved a low detection limit (10 cfu/mL) (Figure 8). The authors then stated that their fast and label-free method can be used in other pathogen and bacterial detection techniques by utilizing the suitable antibodies.

In the same year, Elnaz Akbari et al. fabricated another biosensor using CVD-grown graphene coated with poly-methyl methacrylate to detect *E. coli* bacteria (Figure 7).[22] The performance of the sensor was analytically demonstrated, and the bacteria concentration control parameter was presented. When the bacteria are adsorbed, the changes in the drain-source current and conductance characteristics are plotted. They also simulated other models using an artificial neural network (ANN) and support vector regression (SVR) techniques to assess the performance and the I–V characteristic of the sensor. The results showed that ANN is best suited to predict the sensor performance compared with SVR and that it achieved a low limit of detection (LOD) of nearly 10 cfu/mL.

Meng Liu et al. used a specific non-covalently adsorbed DNAzyme with graphene surfaces to detect various bacteria of interest.[23] DNAzymes are DNA oligonucleotides that can perform certain chemical reactions and achieve highly specific bacterial detection. In comparison with the existing methods, this method can avoid the expensive and labor-intensive requirements of surface functionalization or modification. In this model, the authors developed a fluorescent general circuitry biosensor with a DNAzyme directly attached to colloidal graphene to detect *E. coli* bacteria. The used DNAzyme acts as a fluorescent producer and detection element, whereas the graphene acts as a transducer to convert *E. coli* detection to measurable signals by creating changes in the fluorescence. The authors have experimentally shown that this method can be efficiently utilized for single-cell detection by incorporating a cell culture step of about 10 h for *E. coli*, which is less compared with the conventional methods.

Ebola is a highly dangerous epidemic disease, and its early detection is vital to prevent serious outbreaks. Yantao Chen et al. utilized rGO as a conducting channel to develop an FET sensor for Ebola detection.[24] An $Al_2O_3$ layer (as a passivation layer and top gate oxide), gold nanoparticles (AuNPs; to fix the probes), and anti-Ebola antibodies were immobilized to the channel, as shown in Figure 9. They succeeded in the real-time detection of the Zaire strain of the Ebola virus with a very low detection limit of up to 1 ng/mL; hence, this achievement can definitely be a steppingstone in eradicating Ebola.

Graphene-based sensors were used not only in detecting human diseases but also in environmental, food, agricultural, and aquaculture industries. One such method was reported by Natarajan et al. for detecting the white spot syndrome virus, which is a deadly virus in the aquaculture field.[25] Here, modified GO was used to fabricate the device, where methylene blue and primary and secondary antibodies were anchored to its surface, which resulted in a highly sensitive detection within 35 ± 5 min.

Other researchers utilized polymer matrix composites with nanomaterials for the quantitative and qualitative detection of pathogens. Navakul et al. described a method for the detection of dengue fever by utilizing GO with a polymer composite material and electrochemical impedance.[26] The GO polymer mixture was coated on a gold electrode, and the measurements were performed by plotting the impedance spectra. Here, the GO was used to make the copolymers electrically conductive, thus resulting in the dengue virus detection with a low detection limit of 0.12 plaque-forming units (PFU)/mL.

Additionally, the unique properties of graphene allow microfluidic integration, which provides a platform to merge chemical and biological mechanisms into a single unit. It also provides improvements in portability and real-time detection with a minimum number of samples.[29,51,52] Many researchers also proposed enzymatic microfluidic biosensing systems for developing point-of-care testing.

Singh et al. fabricated a microfluidic rGO-based biosensor for detecting the influenza virus (H1N1).[53] They used three microelectrodes coated with rGO, a gold working electrode, a platinum reference electrode, and a gold counter electrode, which were engineered on a glass substrate along with a SU-8 mold for a microfluidic channel. This highly selective method detected the virus with a low detection limit of 0.5 PFU/mL.

The analysis of the changes in the surface charge density of diseased cells provides relevant information regarding the state of many diseases such as Malaria.[54] Ang et al. described a microfluidic GFET sensor and flow–catch–release sensing for detecting and analyzing single malaria parasites infected cells.[51] When an infected cell binds to the sensor, it creates changes in the drain-source conductivity of the channel, which drops down to nearly 10%.

Aptasensors are aptamer-based biosensors that use aptamers as bioreceptors or transducers. Also, aptamers are oligonucleotides that can bind to a specific molecule with relatively high affinity and specificity. Several graphene-based viral bio-detectors with aptamer-based interactions were proposed for improved performance because they are easy to synthesize and modify and can bind to a broad range of targets.[28] They can also use optical principles, such as fluorescence or electrical properties, to detect molecules.

**6. Hybrid or Advanced Graphene-based Virus biosensors**

Combining graphene with nanoparticles or metallic structures such as those used to increase terahertz coupling[45,55–60] provides novel nanocomposites with more desirable properties. These metallic nanoparticles can be palladium, antimony, silver, or gold. Among these, AuNPs are most widely used with graphene, as they have many unique characteristics, including biocompatibility, low toxicity, and electrical and catalytic features.[61] In addition, graphene-based compounds are also used in SPR biosensors. Also, combining graphene with single-layered transition metal dichalcogenides provides a new platform for developing biosensors that can detect a wide variety of pathogens. Among these, graphene–molybdenum sulfide ($MoS_2$) heterostructures and ZnO–graphene heterostructures have led to new research lines in photon-sensing devices.[60] Sinan Aksimsek et al. proposed a more sensitive biosensor using $MoS_2$–graphene hybrid structures, which showed a considerable increase in sensitivity with more $MoS_2$ and graphene layers.[41]

The rapid detection of the norovirus, which causes food poisoning, was not possible using conventional methods such as ELISA and PCR. Thus, several researchers proposed various biosensing platforms. Chand et al. fabricated a screen-printed carbon electrode in a microfluidic platform,[28] where a carbon electrode was modified with a graphene particle-AuNPs composite (Figure 10), which offered a stable substrate for aptamer immobilization. The norovirus detection was based on aptamer–target interactions, where the aptamer was tagged with ferrocene as a redox probe. When the norovirus binds to the aptamer, an increase in the capacitance of the electrode results in the detection of the virus within a total time of 35 min in the spiked blood sample.

In 2017, Chandan Singh et al. fabricated a microfluidic biosensor for detecting *Salmonella* bacteria by using a GO/carboxylated multiwalled carbon nanotubes composite microelectrode.[62] This wrapping resulted in higher antigen loading, and the method showed superior biosensing characteristics within a detection time of 30 min. Also, graphene with silver nanoparticles or AuNPs was discussed for the ultrasensitive detection of *Salmonella typhimurium*, the hepatitis C virus, and the avian influenza virus (H7) with low detection limits in the order of the picograms per milliliter of the virus.

Lin-Xia Fang fabricated a carbon electrode-based biosensor, where the electrode was modified with AuNPs, graphene, and flower-like VS2 (AuNPs/Gr–VS2/glassy carbon electrode; GCE) and immobilized with aptamer for the detection of the avian influenza (H5N1) virus.[63] The graphene provided a large surface area to hold more AuNPs, which in turn resulted in a highly sensitive virus detection with a low detection level of the order of $5.2 \times 10^{-14}$ M. Fereshteh Chekin et al. researched the integration of GCEs with porous rGO and MoS$_2$, which was immobilized with the human papillomavirus (HPV)-16 L1 specific aptamer for the selective detection of the HPV, and this method resulted in a low detection limit of 0.1 ng.mL$^{-1}$.[64]

Hongyan Fu et al. theoretically found out that the use of graphene on top of the gold films in SPR biosensors greatly increased the sensitivity and showed that the sensitivity of their sensor, which had 20 graphene layers, increased up to 50%.[29] Additionally, Mushfequl Bari et al. examined the change in the surface plasmon resonant frequency (SPRF) and sensitivity with respect to the target concentrations by increasing the number of graphene layers and found that a greater number of layers contribute to higher values in SPRF and sensitivity.[61]

Qi Wang et al. proved the utilization of the large specific surface area of GO in SPR biosensors. The staphylococcal protein A antigen was directly immobilized to the optical fiber sensor surface to develop the ultrasensitive detection of human immunoglobulin G, showing a LOD of nearly 0.5 µg/mL [64].[65] Additionally, Yijia Wang et al. proposed a graphene and gold nanocluster incorporated into a GCE for the detection of human immunodeficiency virus with a detection limit of 30 aM.[66] Moreover, graphene quantum dots were used with gold-embedded polyaniline nanowires by Dutta Chowdhury et al. in their highly sensitive impedimetric biosensor for the detection of the hepatitis E virus.[36]

Many researchers succeeded in developing highly sensitive graphene-based biosensors with picomolar LODs (citations).[67,68] However, a report with actual biological samples is not available. Graphene-based FETs are the most studied biosensors because they can be easily integrated with the commercially available read-out systems, but the leakage current that takes place in them due to the lack of the intrinsic bandgaps in graphene sometimes results in low sensitivity.[45] Solving the leakage issue using proper doping and bandgap engineering techniques will contribute to the efficient implementation of the graphene biosensors.

**7. Future Perspective**

Even in this modern era of technologically-developed diagnostic and medical equipment, viral infection is still a cause of dangerous threats and challenges to public health. Also, appropriate diagnostic equipment for the early and rapid detection of pathogens is still lacking. The devastating rapid spread of COVID-19 around the world has caused severe impacts on the society we live in today and on the world economy, and the need for effective tests to diagnose if people have or had the virus is increasing day by day. The versatility of graphene material has attracted interests from both academia and industry to make sensors for the SARS-CoV-2 viral strain. The use of functionalized graphene[69,70] in PCR sensing, such as in the case of COVID-19, is promising; Recent research shows that when SARS-CoV-2 spike antibody is conjugated to a graphene sheet used as the sensing area (Figure 11), the sensor was able to detect SARS-CoV-2 virus in clinical samples, the SARS-CoV-2 antigen in standard buffer and transport medium; and cultured SARS-CoV-2 virus.[71] Moreover, novel applications such as the integration of GO screen-printed flexible-impedance biosensors in textile, GO screen printed flexible impedance biosensors, with a detection limit of 10 ng/mL, is capable of uncovering exposure to the virus before symptoms manifest, and therefore help in the tracking of outbreak sources[72].

Although graphene-based biosensors are rapidly developing, and a lot of research has been dedicated to this field in recent years, for the successful implementation of these proposed models, several scientific and engineering problems still need to be tackled. Moreover, more research is required to analyze the prolonged use, biocompatibility, and toxicity of this wonder element, especially in the fabrication of wearable and implantable sensors. The continuously changing viral strains require adaptive and easy restructuring equipment. CVD is the most common synthesis method used in the graphene-based biosensor applications, but it still requires more controlled and high-yield synthesis and fabrication methods. Recently, different types of graphene nanosheets and polymer composites have been commercially available, making it easier to fabricate appropriate biosensors. Overall, more research in the utilization of graphene-based materials will surely result in the rapid and early detection of pathogenic organisms, which may require interdisciplinary collaborations from different scientists in the fields of material science, chemistry, physics, electrical engineering, and medicine.

**8. Conclusions**

Viral infections are the main cause of death and serious pandemics throughout the world. The availability of rapid and sensitive mechanisms for the early detection of pathogenic organisms is of great importance in the healthcare field. Graphene is a wonder element with outstanding properties, and its large surface area is an excellent candidate for the fabrication of different types of biosensors. It is also an excellent electrode material, and it has a 2D structure with unique properties. A wide variety of ultrasensitive detectors were discussed in this paper. Yet the clinical application of most of these proposed methods requires more research and testing, and many challenges from the large-scale synthesis of graphene to its biocompatibility issues need to be solved.

**Acknowledgements**

This work was supported by funds from the UAE University Startup project No. 31N312 and the UPAR project No. 31N393. A.E.M and S.P. prepared the draft. M.B. and W.K.P participate in the write-up.

Received: ((will be filled in by the editorial staff))
Revised: ((will be filled in by the editorial staff))
Published online: ((will be filled in by the editorial staff))

**Table 1.** Various sensing modalities, and its salient features. Adapted from [70], 2017, Hindawi

|  | PCR | ELISA | qPCR (Quantitative) |
|---|---|---|---|
| Equipment | standard | standard | fluorescence detection |
| Reagent cost | low | moderate | costly |
| Detection limit (per $\mu$L) | 1–10 ng | 0.01 ng | 0.25 pg |
| Quantitative | not quantitative | semi-quantitative | quantitative |

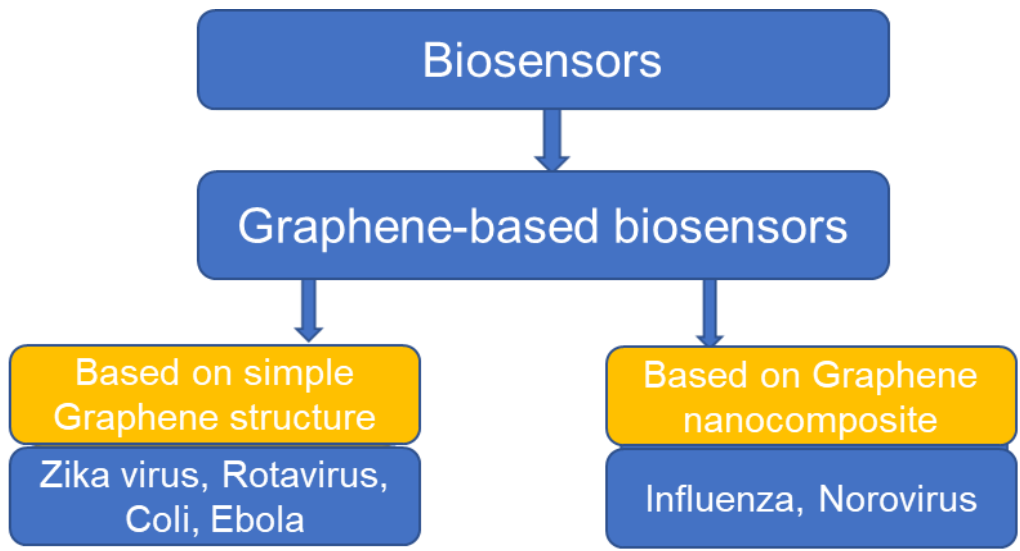

**Figure 1.** An overview of graphene biosensor categories for viral detection.

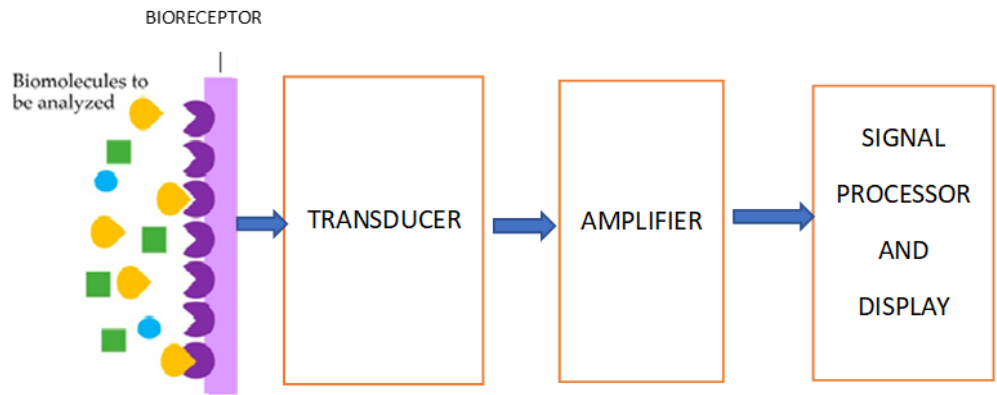

**Figure 2.** A typical architecture of a biosensor system. It consists of mainly four modules bioreceptor, transducer, amplifier and a signal processing unit. Adapted with permission.[11] 2017, MDPI

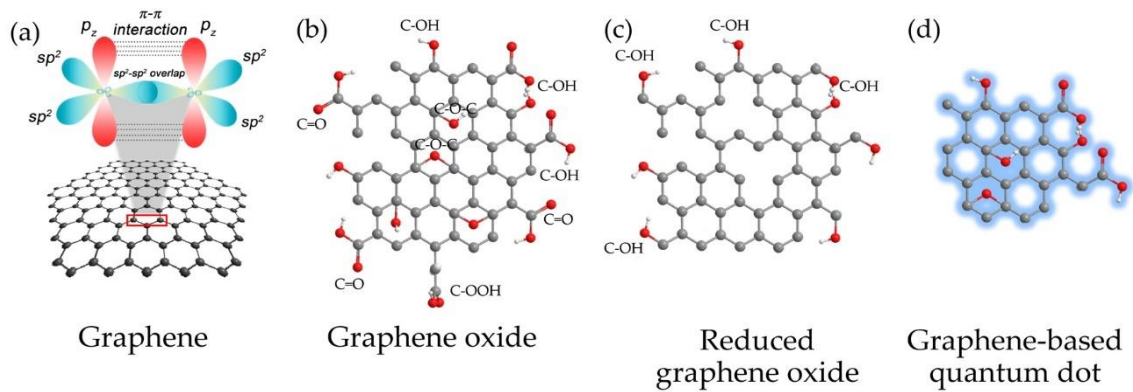

**Figure 3.** Structure of Graphene materials used in the development of biosensor. Reproduced with permission.[11] 2017, MDPI

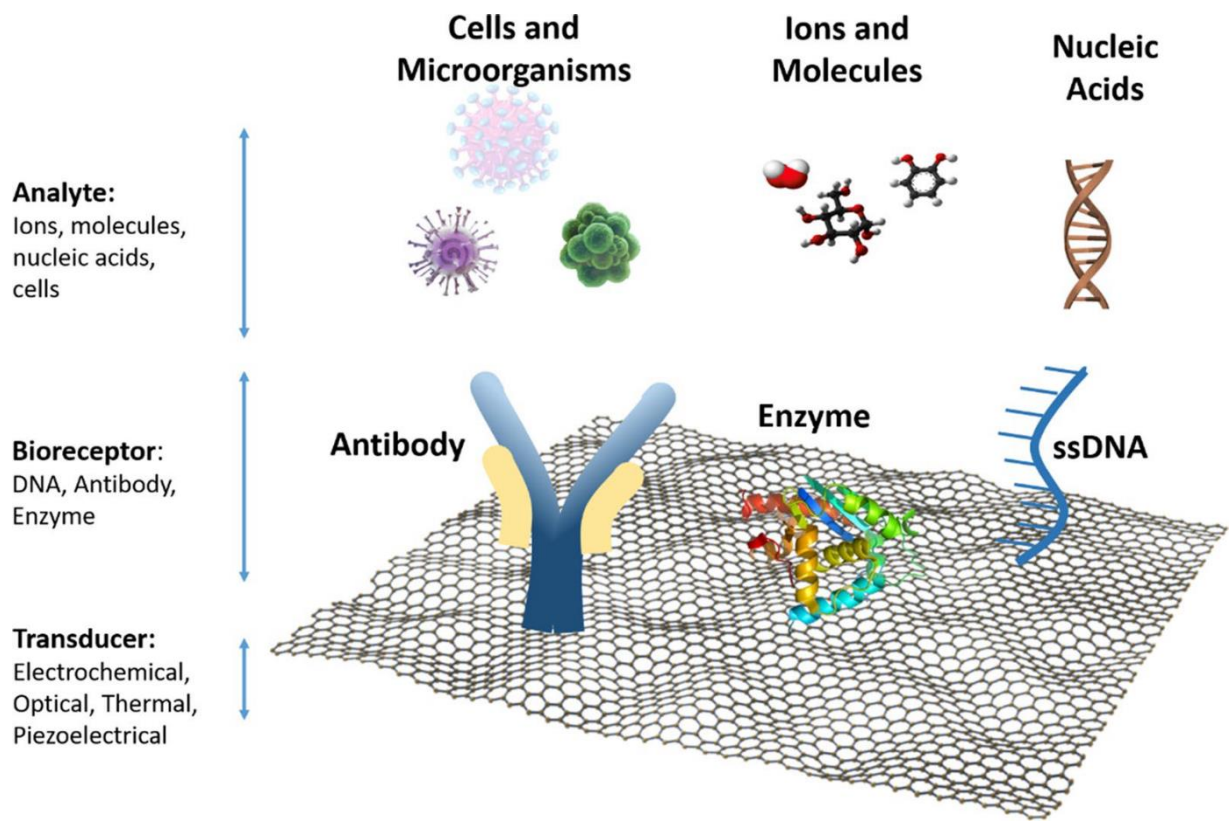

**Figure 4.** A Graphene-based biosensors platform and its components. Reproduced with permission.[73] 2018, BMC

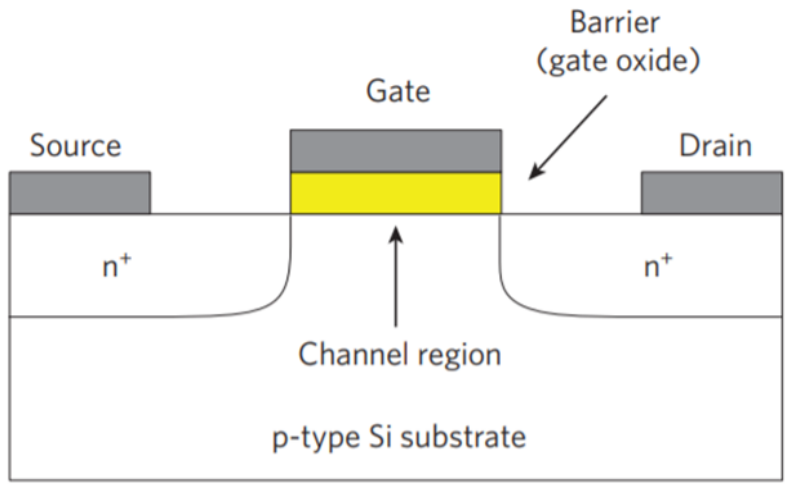

**Figure 5.** A cross-section of a GFET. The channel region is a thin depletion layer under the gate oxide.

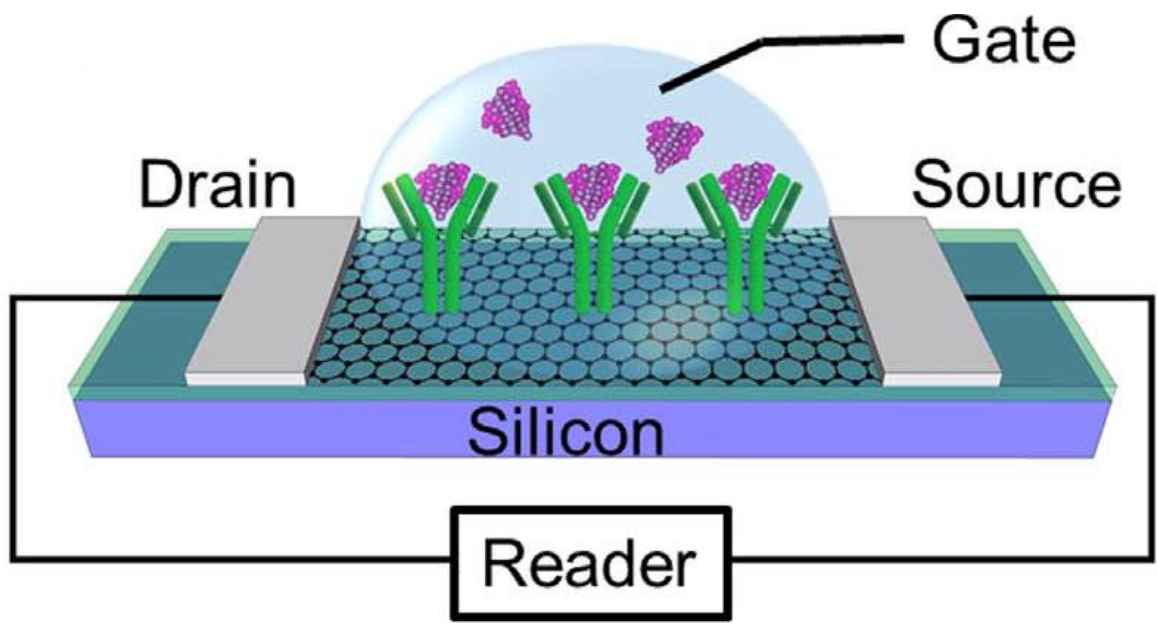

**Figure 6.** A graphene channel biosensor for zika virus detection. Adapted with permission.[49] 2018, Elsevier

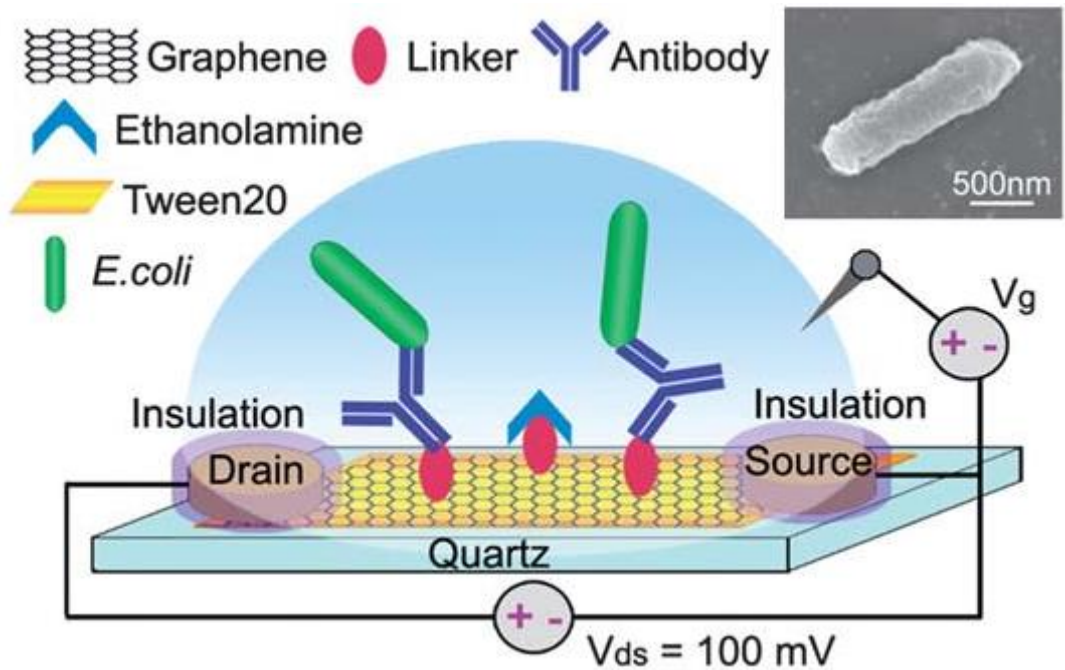

**Figure 7.** Graphene FET for detection of E. coli bacteria. Inset: Scanning electron microscopy (SEM) image of an E. coli on antibody functionalized graphene. Reproduced with permission.[21] 2011, Royal Society of Chemistry

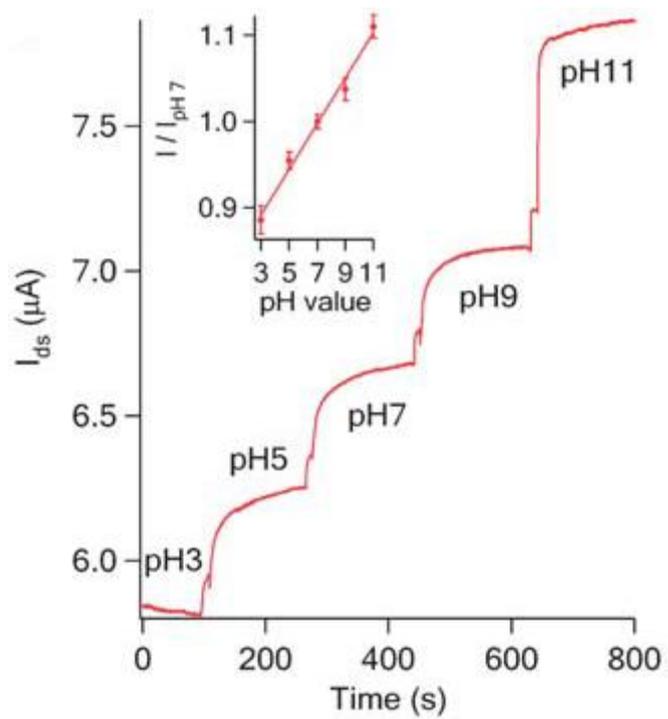

**Figure 8.** Real-time current with time-varying concentration solutions for different pH values.[21] 2011, Royal Society of Chemistry

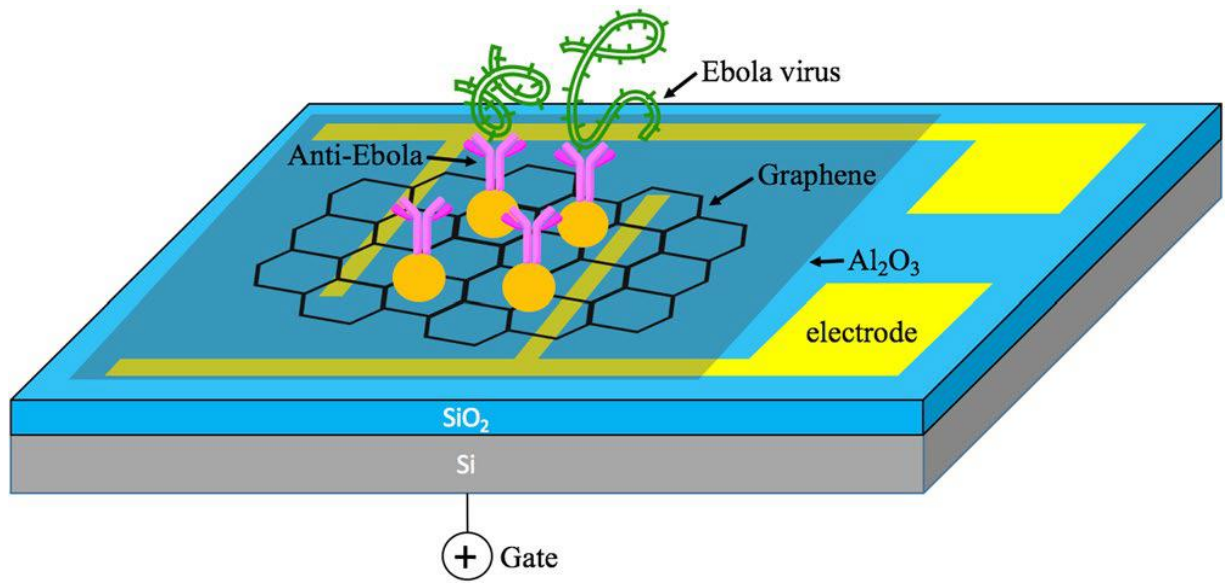

**Figure 9.** Schematic diagram of the rGO-based FET for detecting Ebola virus. Reproduced with permission.[24] 2017, Nature Research

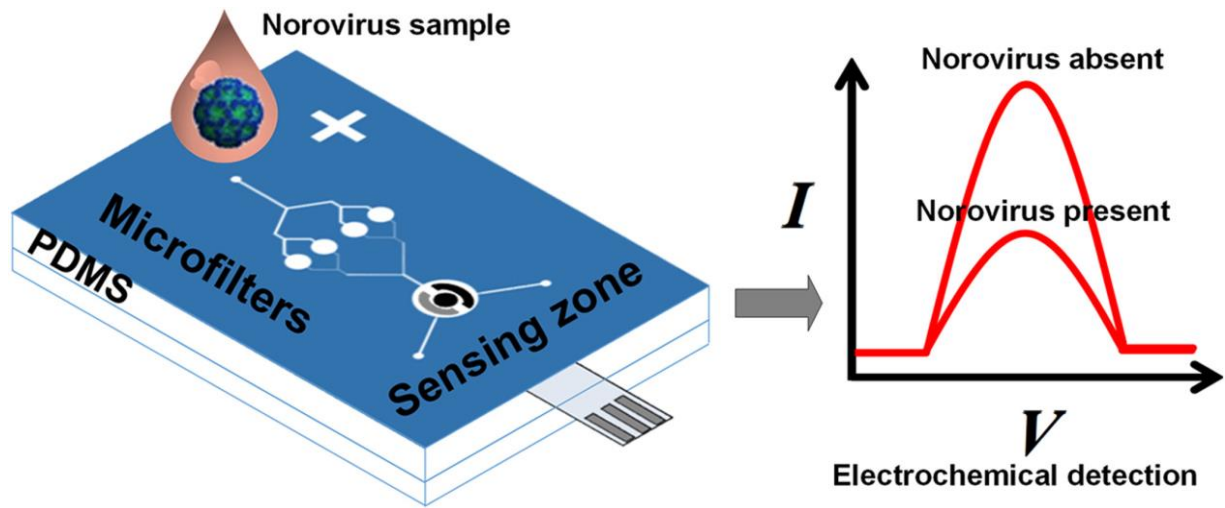

**Figure 10.** Microfluidic carbon electrode with Grp-AuNPs composite for Norovirus detection. Reproduced with permission.[28] 2017, Elsevier

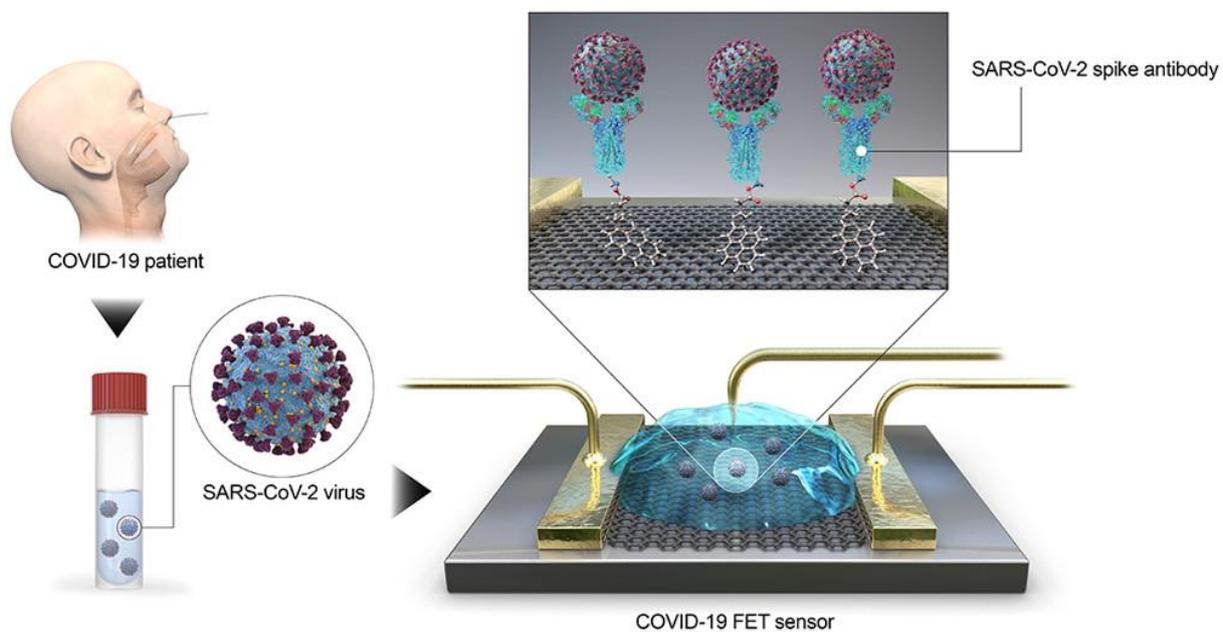

**Figure 11.** Schematic diagram of COVID-19 FET sensor operation procedure. Graphene as a sensing material is selected, and SARS-CoV-2 spike antibody is conjugated onto the graphene sheet via an interfacing molecule as a probe linker. Reproduced with permission.[71] 2020, ACS publications

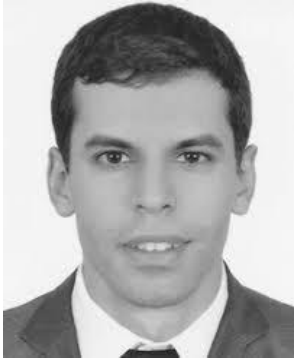

Prof. Amine El Moutaouakil received the Ingenieur d'Etat in electromechanical engineering from Ecole des Mines, Morocco, in 2005, and the M.E. and Ph.D. degrees from Tohoku University, Japan, in 2008 and 2011, respectively. He was a Research Staff Member with the Nippon Telegraph and Telephone Corporation, Japan, where he was involved in InP-based HEMTs and MMICs. Since 2017, he has been with the UAE University, United Arab Emirates. His current research interests include the design, fabrication, and characterization of electronic devices for applications in ultrahigh and terahertz frequencies, and the application of semiconductor materials in security, biomedical and communication applications.

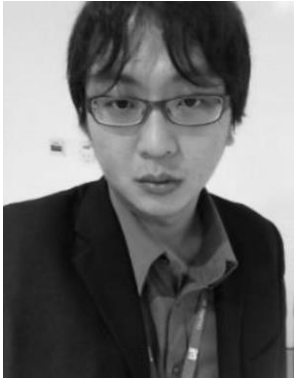

Dr. Weng Kung is a Research Group Leader and Principal Investigator at International Iberian Nanotechnology Laboratory (INL). His research group interest focuses on developing and translating technological innovations (e.g., microscale nuclear magnetic resonance, electron spin resonance) for the next generation of molecular phenotyping in precision medicine. Prior to joining INL, he was Research Scientist (2014-2015) and Junior Investigator (2009-2013) at BioSyM-SMART, Massachusetts Institute of Technology (MIT). BioSyM-SMART Centre is the first research institution established outside Boston, under the joint venture between MIT and National Research Foundation of Singapore.